\documentclass[11pt,a4paper]{article}

\usepackage{pdfpages}

\usepackage{graphicx}
\usepackage[sort&compress,square,numbers]{natbib}
\usepackage{amsmath}
\usepackage{times}

\usepackage[left=1.1in,right=1.1in,top=1in,bottom=1in,head=0.4in]{geometry}
\begin{document}



**Why is combinatorial communication rare in the natural world, and why is language an exception to this trend?**

(short title: **Evolution of combinatorial communication**)

Thomas C. Scott-Phillips [1,*]

Richard A. Blythe [2,*]

[1] Evolutionary Anthropology Research Group, Department of Anthropology, Durham University, Dawson Building, South Road, Durham, DH1 3LE, UK

[2] SUPA, School of Physics and Astronomy, University of Edinburgh, James Clerk Maxwell Building, Mayfield Road, Edinburgh EH9 3JZ, UK

[*] The two authors contributed equally to this work

1

**Abstract**

In a combinatorial communication system, some signals consist of the combinations of other signals. Such systems are more efficient than equivalent, non-combinatorial systems, yet despite this they are rare in nature. Why? Previous explanations have focused on the adaptive limits of combinatorial communication, or on its purported cognitive difficulties, but neither of these explains the full distribution of combinatorial communication in the natural world. Here we present a nonlinear dynamical model of the emergence of combinatorial communication that, unlike previous models, considers how initially non-communicative behaviour evolves to take on a communicative function. We derive three basic principles about the emergence of combinatorial communication. We hence show that the interdependence of signals and responses places significant constraints on the historical pathways by which combinatorial signals might emerge, to the extent that anything other than the most simple form of combinatorial communication is extremely unlikely. We also argue that these constraints can be bypassed if individuals have the socio-cognitive capacity to engage in ostensive communication. Humans, but probably no other species, have this ability. This may explain why language, which is massively combinatorial, is such an extreme exception to nature's general trend for non-combinatorial communication.

**Keywords**: animal signals, combinatorial communication, communication, evolution, language, ostension

# 1. Introduction

In a combinatorial communication system, some signals consist of the combinations of other existing signals. The most basic version involves two individual signals that are combined to refer to something that is not simply the amalgamation of whatever the two individual signals refer to, but something different. In other words, combinatorial communication includes at least one *composite* signal, whereas in non-combinatorial communication, all the signals are *holistic* (figure 1). For example, putty-nosed monkeys are reported to have two distinct alarm calls, one for each of two predators: leopards (a 'pyow' sound) and eagles (a 'hack' sound) [1,2]. When one or the other of these calls is produced on its own, the monkeys take appropriate evasive action: climbing up and into the trees for leopards; climbing down and into the bushes for eagles. However when the two calls are produced together ('pyow-hack'), the effect is not the simple combination of these i.e. the monkeys do not behave as if avoiding both types of predator. Instead, the call seems to presage the movement of the group to a new location (perhaps e.g. because there is a shortage of fresh food at the present location).

[figure 1 about here]

Combinatorial communication has one obvious adaptive advantage over equivalent non-combinatorial systems: fewer elements are required to express the same number of possible messages, and so it allows information to be transferred more efficiently than in a system in which each signal has a distinct form [3]. Despite this potential advantage it is rare in nature, and where it does exist it is, with one salient exception, simple and limited [4,5]. Many systems (e.g. honeybee dance) have a signal which has an effect that is equal to the sum of its component parts, but very few have a signal whose effect is different to this sum (in the terms of figure 1, $Z = X + Y$ is common, but $Z \neq X + Y$ is not). The salient exception is human language, which is massively combinatorial. Indeed, there are multiple different types of combination that can, along with other linguistic phenomena (syntax, phonetics, etc.), contribute to meaning. Otherwise the only well-

attested examples are the putty-nosed monkey case described above, and the quorum sensing communication of bacteria [6].

Why, then, is combinatorial communication so rare, and why is language such an extreme exception? One possible explanation might be that combinatorial communication is cognitively challenging in some way [7,8]. However, why this should be the case is unclear: there is no particular reason to think that signal combinations should be difficult to process. On the contrary, the bacterial data suggests that it does not pose any significant cognitive challenges [6]. Other previous explanations have focused on the limits of the adaptive benefits of combinatorial communication. One analysis shows that the benefits associated with combinatorial communication are only realized when the total number of signals in the system exceeds a threshold level [3,9]. However, this prediction is not supported by the data: several systems have many more signals than this threshold level, but still do not combine them (non-human primate gestural communication, for example [10]); and the putty-nosed monkey system described above has fewer signals than this threshold level, but is combinatorial nevertheless. Finally, one other analysis concludes that combinatorial systems are more susceptible to dishonesty than non-combinatorial systems [11]. However, this does not explain why language should be such a clear exception.

In this paper we develop a new explanation of the distribution of combinatorial communication in nature. The historical pathways by which traits evolve are an important source of constraint on biological form [12], yet most models of the evolution of communication, including all those focused on combinatorial communication, ask only how one type of communicative system can evolve from another, and not how an initially non-communicative behaviour can take on a communicative function. Specifically, previous models do not include strategies for signaller and receiver prior to the emergence of a signal, despite the fact that such prior strategies are likely to have considerable impact on the form of the eventual communication system [13,14]. We present a deterministic, nonlinear dynamical model that includes such strategies. We use this to derive a

number of principles regarding the origins of composite signals, and hence show that the emergence of composite signals is subject to significant historical constraints, to the extent that even the most simple forms of combinatorial communication are likely to be uncommon, and anything more complex vanishingly so. We then argue that these constraints can be bypassed if a species has the social cognitive abilities to communicate ostensively i.e. in a way that involves the expression and recognition of communicative and informative intentions. Humans, but probably no other species, have this ability.

## 2. The emergence of communication

Before we present the model, it is necessary to briefly describe the two classic ways in which communication systems can emerge. In general, new signals emerge through either ritualisation or sensory manipulation (also called sensory exploitation) [14-16]. In ritualisation, previously existing cues are exapted for use as signals (a cue is a behaviour that is informative for other organisms, but was not selected to be so [15,17]). For example, the use of urine to mark territorial boundaries probably first began when animals urinated simply through fear, when they were at the edge of familiar territory. This acted as a cue to other animals, who make use of that information. This in turn provides a selection pressure on the focal organism to urinate when and if it wants/needs to inform others about the range of its territory. In sensory manipulation, previously existing behaviour is exapted for use as a response. For example, male scorpionflies capture large prey and then offer them to females who feed on them during copulation [18]. The offering of prey by the male probably initially evolved because the female had a pre-existing mechanism that prioritized the opportunity to feed on large prey, and so the presentation of food gave the male an opportunity to mate. There was then later positive selection on the female to accept the prey in exchange for copulation [15].

These processes, ritualisation and sensory manipulation, constrain the form that new signals can take. Our previous model showed that if a particular behaviour does not exist for reasons independent of communication prior to either of these processes occurring, then it cannot become a signal or response, regardless of its adaptive value for signaller or receiver [14]. Prospective signals and responses must already provide fitness benefits to either receiver or signaller, respectively, if they are to actually evolve into signals/responses. In other words, there must be some sort of trigger, external to the (proto-)communicative interaction (i.e. either a cue or a coercive behaviour), to cause a signal to actually emerge. With this background in place, we now develop a formal model of the emergence of specifically combinatorial communication systems.

### 3. Basic set up of model, and classification of communication systems

The environment can be in any one of a number of different states, $E \in \mathscr{E}$. Given this state, one agent (the actor) performs an action, $A(E) \in \mathscr{A}$. Another agent (the reactor) then performs a reaction, $R(A) \in \mathscr{R}$. Payoffs are determined by the different combinations of states and reactions. This much is the same as the standard game-theoretic approach to modelling communication. However we differ from the standard approach in our specification of the sets $\mathscr{E}$, $\mathscr{A}$, and $\mathscr{R}$. In particular, in our model these sets include default settings ($E_0$, $A_0$ and $R_0$), which correspond to the agents doing nothing, and which are orthogonal to all other members of the set. Typically, these default settings are not included in game-theoretic models of communication. However, they are critical to understanding how communication can emerge from a state of non-communication [14].

With the exception of the defaults $E_0$ and $A_0$, environments and actions can be combined with other environments and actions. We denote these composites as $E_i \circ E_j$ and $A_i \circ A_j$ respectively. These composites are, like their component parts, members of $\mathscr{E}$ and $\mathscr{A}$ respectively,

and composition is commutative (i.e. order does not matter, so $E_i \circ E_j = E_j \circ E_i$ and $A_i \circ A_j = A_j \circ A_i$).

$A(E)$ and $R(A)$ are deterministic functions of $E$ and $A$ respectively. Together they comprise an agent's strategy. If there are non-composite environments $E_i$ and $E_j$ in which the actions $A_i$ and $A_j$ are performed, then the composite action $A_i \circ A_j$ is performed in the composite environment $E_i \circ E_j$. Two agents are members of the same species if they share the same functions $A$ and $R$ for all possible environments and actions respectively (i.e. two individuals $i$ and $j$ are members of the same species $\Leftrightarrow A_i(E) = A_j(E) \ \forall \ E$ and $R_i(A) = R_j(A) \ \forall \ A$).

A signal is any non-default action which yields a non-default reaction, and any such reaction is called a response. A communication system is a set of more than one signal-response pair. Within a communication system, each given pair of actions $A_i$ and $A_j$ (where both $\neq A_0$) can be classified in one of three ways:

- *non-composite*: a pair is non-composite if the composite of the two actions is produced only in composite environments, and it in turn yields the default reaction i.e. there is no $E_k \neq E_i \circ E_j$ such that $A(E_k) = A_i \circ A_j$, and $R(A_i \circ A_j) = R_0$.
- *pseudo-composite*: a pair is pseudo-composite if the composite of the two actions is produced only in composite environments, and it in turn yields a non-default reaction i.e. there is no $E_k \neq E_i \circ E_j$ such that $A(E_k) = A_i \circ A_j$, while at the same time $R(A_i \circ A_j) \neq R_0$.
- *fully-composite*: a pair is fully-composite if the composite of the two actions is produced in at least one non-composite environment, and it in turn yields a non-default reaction i.e. $\exists \ E_k \neq E_i \circ E_j$ such that $A(E_k) = A_i \circ A_j$, and $R(A_i \circ A_j) \neq R_0$. A combinatorial communication system is a system that includes at least one pair of fully-composite actions.

(Logically, there is a fourth possible class, where the composite of two actions is produced in at least one non-composite environment, and it in turn yields the default reaction. However, such a pair is unstable, since the default reaction produces only a zero payoff for the actor, by definition,

7

and this is outweighed by the maintenance and production costs of the composite pair. We therefore ignore this possibility in the subsequent analysis.)

Communication is vulnerable to an instability caused by dishonesty. How communication systems remain stable in the face of this problem is an important and much studied question for the evolution of communication [15,19]. Here, however, we are concerned with a different question, namely: assuming that communication is evolutionarily stable, what are the different evolutionary pathways by which (combinatorial) communication can emerge? We thus wish to avoid issues of stability, which might complicate our analysis, and so we assume that at least one of the different mechanisms that can stabilise communication is in place. In particular, we find that ascribing a direct benefit for successful communication to both signaller and receiver [as in e.g. 20,21], or imposing kin discrimination (i.e. agents can observe the actions of their conspecifics only), leads to the same set of dynamical equations (see *Supplementary Information*). We expect other possible mechanisms to lead to the same or similar principles for the emergence of combinatorial communication as those we set out below.

### 4. Dynamics

We now derive the dynamics for the model. The basic principles of the model are that: the frequency of a particular environment is given by $f(E)$; if an agent performs a reaction $R$ in environment $E$, there is a payoff $s(R|E)$; the actor must also receive a payoff, either directly or indirectly, for communication to be stable (see above); and there is a cost associated with each agent's strategy i.e. the cost of having the capacity to behave in a non-default way in the first place (this is $\chi$ for actions and $\eta$ for reactions; see below). This is distinct from the cost associated with each individual behaviour, which we include as part of the payoff $s(R|E)$. These two types of cost can be thought of as *maintenance costs* and *energy costs* respectively, and the inclusion of both is an

important difference between our model and previous models of the emergence of communication [e.g. 20,21].

We first write down the dynamical equations for the frequencies of the various strategies in the population, and we define $\psi(A,E)$ as the fraction of agents who perform action $A$ in environment $E$, and $\phi(R,A)$ as the fraction of agents who perform reaction $R$ in response to action $A$. As we show in the *Supplementary Information*, we have:

$$\frac{d}{dt}\psi(A,E) = \psi(A,E)\left[u(A,E) - \sum_{A'} u(A',E)\psi(A',E)\right]$$

where

$$u(A,E) = f(E)\sum_{R} s(R|E)\phi(R,A) + \sum_{E' \neq E, E_0} f(E \circ E')\sum_{R} s(R|E \circ E')\sum_{A'} \psi(A',E)\phi(R, A \circ A') + \chi\delta(A, A_0) \quad (1)$$

($\delta$ is the Kronecker delta symbol, which equals 1 if the two arguments are identical, and 0 otherwise.) The equation has the same structure as the replicator equation in standard evolutionary game theory [22], in which $u(A,E)$ is the fitness of the rule $E \rightarrow A$. There are three contributions to this fitness:

- The first term gives the fitness of the rule $E \rightarrow A$ in the environment $E$, given the current distribution of the possible rules $R(A)$.
- The second term gives the net fitness of the rule $E \rightarrow A$ in all the composite environments that include $E$, given the current distribution of the possible rules $R(A)$. This term is unique to composite signals. It has the consequence that the spontaneous emergence of composite signalling strategies (either actions or reactions) can be favourable.
- The third term accounts for the cost, $\chi$, of having a mechanism for producing non-default actions.

Similarly, for the frequency $\phi(R,A)$ of the rule $A \rightarrow R$, we get:

$$\frac{d}{dt}\phi(R,A) = \phi(R,A)\left[v(R,A) - \sum_{R'} v(R',A)\psi(R',A)\right]$$

where

$$v(R,A) = \sum_{E} f(E)s(R|E)\psi(A,E) + \sum_{\langle E_1,E_2\rangle} f(E_1 \circ E_2)s(R|E_1 \circ E_2)\sum_{A_1,A_2} \psi(A_1,E_1)\psi(A_2,E_2)\delta(A_1 \circ A_2, A) + \eta\delta(R,R_0)$$

(2)

As above, there are three terms to this equation, which correspond to: the fitness of the rule $A \to R$ in non-composite environments; the average fitness of a composite action $A = A_1 \circ A_2$ performed in composite environments; and the cost, $\eta$, of having a mechanism for producing non-default reactions. Also as above, we derive this equation step-by-step in the *Supplementary Information*.

These equations can be applied to any specific set of components i.e. states of the environment; possible actions and reactions; and parameters. In the *Supplementary Information* we define these components for a specific model, in order to test the general predictions we derive below. Its results are entirely consistent with the principles we set out below.

**5.    Three basic principles for the emergence of combinatorial communication systems**

We now derive three basic principles that govern how a combinatorial communication system might emerge. We are interested, primarily, in the case where two actions, $A_i$ and $A_j$, are composed to become the action that is used to signal an elementary environmental state $E_k$ that is unrelated to $E_i$ and $E_j$. We will not discuss higher-order composite actions (i.e. those where one or both components are themselves composite actions), but we have no reason to think that the general principles we derive here should be any different in that case.

*Principle 1: Only homogeneous populations are evolutionarily stable.*

For any given environmental state, the fitness, $u(A,E)$, of each action rule is independent of the frequencies, $\psi(A',E)$, of its competitors. The action rule with the highest fitness will then grow at the expense of all its competitors until it is the only one remaining. If two or more rules have the

same fitness as each other, drift will likewise eliminate all but one of them. These observations also applies to reaction rules. Hence, only homogeneous populations, in which all agents have the same set of rules, are evolutionarily stable. We thus assume homogenous populations in what follows.

*Principle 2: New non-composite signals cannot emerge without an external trigger.*

Given some pre-existing signalling system, a new non-composite signal is the action, $A$, in the set of rules $E \rightarrow A \rightarrow R$ ($A \neq A_0$, $R \neq R_0$), where: (i) there is currently no environment in which the new action $A$ is performed; (ii) $A$ is not a composite of any two existing actions in the system and (iii) the existing reaction to $A$, produced by itself or in combination with any other action $A'$, is the default reaction. In a homogeneous population (see principle 1), condition (i) implies that the first term in equation (2) vanishes for both $v(R,A)$ and $v(R_0,A)$ because $\psi(A,E) = 0$ for all $E$; condition (ii) implies that the second term in equation (2) also vanishes for both $v(R,A)$ and $v(R_0,A)$, because $\delta(A_i \circ A_j, A) = 0$ for all $A_i$, $A_j$; and condition (iii) implies that the first two terms of equation (1) are the same for both $u(A,E)$ and $u(A_0,E)$ because $\phi(R',A) = \phi(R', A \circ A') = \delta(R',R_0)$ for all $A'$. Consequently, for any new non-composite signalling behaviour, we always have that:

$u(A,E) - u(A_0,E) = -\chi$ in any environment

$v(R,A) - v(R_0,A) = -\eta$

In other words, adding either the action or reaction component of this signalling behaviour carries a cost for every individual in the population. Hence a completely new signalling behaviour, that involves an action that is not currently part of the signalling system, cannot emerge without some external trigger as described in §2. This result is an extension of our previous result, that a signal cannot be added to an existing state of non-communication without an external trigger (cue or coercion) [14]. The observation here is that this issue also applies to the addition of a any non-composite signal to an existing communication system. This point will be important when we discuss human linguistic communication, in §7, below.

*Principle 3: New composite signals can emerge without an external trigger.*

Consider a pair of existing signals, $A_i$ and $A_j$. If there is an environment where the combination of these two signals would provide a cue (useful information) for other organisms, then their co-production can lead to the evolution of a composite signal. More formally, if the environments $E_i$ and $E_j$ trigger the actions $A_i$ and $A_j$, respectively, then the composite environment $E_i \circ E_j$ will trigger the composite action $A_i \circ A_j$. Repeating the analysis from principle (2), we find, again, that adding a new action will be costly for all individuals in the population i.e. that $u(A_i \circ A_j, E_k) - u(A_0, E_i \circ E_j) = -\chi$ for all $E$, as before. As with principle (2), this prevents the emergence of a composite signal by sensory manipulation. However unlike principle (2), the emergence of a composite signal by ritualization is possible. That is, there may be a reaction that is not yet an existing reaction to either $A_i$ or $A_j$, and which would be beneficial for the receiver to perform in the composite environment i.e. it is possible for $v(R,A) - v(R_0,A) > 0$ for some $R$. If this condition is satisfied, then, assuming that the relevant environments occur sufficiently often for evolution to occur, we should expect the corresponding reaction to evolve. We will then have arrived at a pseudo-composite signal: $(E_i \circ E_j) \rightarrow (A_i \circ A_j) \rightarrow R \neq R_0$. The key observation here is that the co-production of two existing signals can itself be the trigger required for a new signal to emerge. This possibility is absent in the case of non-composite signals. Note that from here, it is then possible for a new rule $E_k \neq (E_i \circ E_j) \rightarrow (A_i \circ A_j)$ to emerge by sensory manipulation, giving us a fully-composite signal. A concrete demonstration of this possibility is given in the *Supplementary Information*, where we apply our model to the specific case of putty-nosed monkey alarm calls. Note, however, that the emergence of a full-composite signal is not guaranteed (it is possible, for example, that reactions to higher-order compositions could prevent this).

## 6. Why combinatorial communication should be rare

What do these principles imply for the emergence of a combinatorial communication system? One immediate observation is that they explain how fully-composite signals can emerge even in a simple world of just two existing signals. This is noteworthy because it is contrary to a previous analysis, which argued that composite signals should only emerge within systems of multiple (>5) signals [3]. However, the present empirical data suggest, consistent with our analysis, that composite signals can exist even in very simple systems (see e.g. the putty-nosed monkey system described in §1).

Another observation might be that the three principles derived above seem to imply that composite signals should be far more common than non-composite signals. After all, composite signals can emerge without an external trigger (principle 3), but non-composite signals cannot (principle 2). However, this reading fails to take account of the conditions attached to each of these possibilities. There are two in particular that we wish to highlight.

The first is the relative frequency by which the various triggers of the emergence of composite and non-composite signals occur. In particular, although the emergence of a composite signal does not require a trigger external to the system itself, it does require one from within the system. This condition is sufficiently stringent that the emergence of a composite signal is in fact less likely to occur than the external triggers that are required for the emergence of non-composite signals. Here is why. The internal trigger required for the emergence of a new composite signal is a very specific one: that, given the co-production of two existing signals, $A_i$ and $A_j$, then there is a reaction $R$ that: (i) if it were performed in the composite environmental state $E_i \circ E_j$, it would be beneficial to the receiver; and (ii) that this reaction is not yet an existing reaction to either $A_i$ or $A_j$ (see principle 3). In other, more information-centric terms, what is required is that the co-production of two existing signals must be informative about some aspect of the world, beyond what can be deduced from the meanings of the individual signals themselves - and there is no particular reason why this should be the case. In contrast, a new non-composite signal can emerge from *any*

behaviour that an individual might perform (see §2). In other words, there is one specific way that any new signal might be composite, but a vast number of ways, limited only by the number of behaviours the organism can actually perform, that any new signal might be non-composite. Consequently, composite signals should be rare. This is not to say that they cannot emerge, only that their emergence is dependent upon unlikely prior circumstances. Hence, they should be rare, at least in comparison with non-composite signals.

The second condition attached to the emergence of fully-composite signals is that it is dependent upon the instability of other possible systems. Consider a basic system of $E_i \rightarrow A_i \rightarrow R_i$ and $E_j \rightarrow A_j \rightarrow R_j$ (i.e. just the first two signals in figure 1). Principle 3 states that it is then possible for a fully-composite signal $E_k \rightarrow (A_i \circ A_j) \rightarrow R_k$ to emerge, to form the system described in figure 1, without an external trigger. However, it turns out that this is only true if the alternative system, of the basic system plus a holistic signal $E_k \rightarrow A_k \rightarrow R_k$ is unstable. Here is why. In order for the process described in principle 3 to occur, the basic system must be unstable to the addition of $(A_i \circ A_j) \rightarrow R_k$. A necessary condition for this to be the case is that in this system, $v(R_k, A_i \circ A_j) > v(R_0, A_i \circ A_j)$. Using the equation for $v(R,A)$ in §4, we find that for this instability to be present, we require:

$$f(E_i \circ E_j) [s(R_k \mid E_i \circ E_j) - s(R_0 \mid E_i \circ E_j)] - \eta > 0 \qquad (3)$$

However, using the same equation for $v(R,A)$, but now applied to the alternative system (i.e. the one that includes $E_k \rightarrow A_k \rightarrow R_k$ rather than $E_k \rightarrow (A_i \circ A_j) \rightarrow R_k$) we find that for this alternative system to be stable we require:

$$f(E_i \circ E_j) [s(R_k \mid E_i \circ E_j) - s(R_0 \mid E_i \circ E_j)] - \eta < 0 \qquad (4)$$

The equations (3) and (4) clearly contradict each other. This shows that the conditions required for the process described in principle 3 to occur include that the alternative system is evolutionarily unstable (note that this is true whether or not the fully-compositional system described in figure 1 is evolutionarily stable). As such, this is an additional criterion on the emergence of composite signals, and hence on the emergence of combinatorial communication.

In sum, there are at least two conditions that can work to restrict the emergence of composite signals. The first is that the triggers required for the emergence of composite signals are less likely to occur than are the triggers for non-composite signals. The second is that the process of emergence without an external trigger depends upon the instability of any alternative, holistic system. Both these conditions are the consequence of the interdependence of signals and responses, and they help to explain why combinatorial communication is rare in nature.

## 7. Human linguistic communication

There is, of course, one extreme exception to the norm of non-combinatorial communication: human linguistic communication. Here, meaningless sounds (phonemes) are combined into meaningful units (morphemes), which are in turn combined into utterances, whose meaning is a function not only of the morphemes involved, but also the order in which they are combined [a feature called *duality of patterning*: 23,24]. This combinatorial richness gives language its expressive power [5,25]. How can we explain why language is such a clear exception to the general trend for non-combinatorial systems? In this section we use the conclusions from our model to pinpoint and articulate an important difference between human and animal communication. We hence argue that human linguistic communication is simply not subject to the various historical contingencies described in the previous sections - and consequently, combinatorial communication is free to emerge wherever it may be useful.

Human communication depends, at bottom, upon mechanisms of metapsychology: that is, the ability to reason about others' reasons, intentions, beliefs, and so on. Communication of this sort is called *ostensive communication* [26]. Linguistic communication is an instance of ostensive communication that has been made expressively powerful by the development of rich suite of communicative conventions that allow it to be used far more precisely and expressively than it otherwise would [27]. A signaller can, for example, ostensively point to any of the objects in this

room, but with language she can refer to any object in the world. She also can make a request of others by, for example ostensively pushing unchopped vegetables, and a knife, in their direction, but with language she can make requests about things remote in time and space. Other examples are not hard to imagine. In contrast, most, and perhaps all, animal communication depends upon mechanisms of association: causal relationships between stimuli and responses (but see below). Communication of this sort is called *coded communication* [see refs. 26-28 for discussion of the difference between coded and ostensive communication].

Our model in this paper has been a model of coded communication: we have studied how states of the world become *associated* with certain actions, and how these actions in turn become *associated* with certain reactions. Indeed, all models of animal communication that study the emergence of such associations are code models. However, such models do not capture an important fact about ostensive communication: that meaning is not deduced or calculated, even probabilistically, on the back of associations (be they between signal and meaning, or perhaps between signals, context, and meaning), but rather it is inferred, based on the receiver's beliefs about the signaller's intentions [26]. This inference is, unlike the associations that make coded communication possible, made possible by metapsychology [28].

One consequence of this difference is that human ostensive communication, including linguistic communication, is inherently prone to ambiguity. This is generally seen as a defective quality, because it can, on occasion, lead to misunderstanding and other failures of communication. However, it also allows communication to be used in flexible, creative, and open-ended ways - and these ways include the combination of already existing signals. One consequence of this is that the spaces of possible signal forms and signal meanings become continuous, rather than discrete. This development is possible only because signaller have the metapsychological abilities to create the right sort of signal to express their intended meaning, whatever it might be, and because receivers have similar abilities to infer those intended meanings.

Here is an example. Homesigners are deaf children born to hearing parents. Lacking the input of a conventional sign language, they must create new communication systems themselves, and this includes the combination of existing signals [29]. Here is one very simple case [*ibid.*]. The child, Karen, is already familiar with pointing, and also with a 'twist' gesture that means 'open'. She then uses these two behaviours together: she points to a jar of soap bubbles and then, without pausing, produces an iconic 'twist' action with her hands. In doing so, she indicates to the adult that she would like her to open the jar. At first blush this seems unremarkable, but that is only because, as fluent users of ostensive communication, we are fully accustomed to such acts of creation as an everyday occurrence. The point here is not that we can combine things together. It is rather that, because she has the required metapsychological abilities, it is possible for Karen to provide just the right sort of evidence, given her intended meaning and her intended audience. This is ostensive communication. It just happens that in this case the right sort of evidence happens to involve the combination of two existing signals.

Note that in a different context, the meaning of Karen's behaviour could be very different indeed. Suppose, for example, that the adult had just tried to open the jar by twisting it, but had failed, and that this had amused Karen. Now Karen could use the same combined signal to make a humorous reference to this past event. This flexibility is possible only because both Karen and the adult have the metapsychological abilities required. On the ostensive side, Karen produced the signals in such a way that it was apparent that they are in fact one signal, comprised of two parts; this is why she does not pause between the two. On the inferential side, the adult must assess what Karen's intended meaning was, given her knowledge of the context, and of the meanings of the two component parts.

Here is how this example relates to our model. Karen has an existing set of actions that she produces in particular environments, and these receive particular responses from the adult. Specifically:

$E_1$ = Karen wants to refer to an out-of-reach object

$A_1$ = pointing

$R_1$ = attention is focused in the direction of the point

$E_2$ = Karen wants the adult to open something

$A_2$ = 'twist' gesture

$R_2$ = the adult opens the object of mutual attention.

Karen finds herself in a new environment: $E_3$ = Karen wants the adult to open an out-of-reach object. Note that this environment is *not* the sum of the other two: $E_3 \neq E_1 \circ E_2$. Instead, the composite environment $E_1 \circ E_2$ is the co-occurrence of (i) an object that is out-of-reach object; and (ii) an object that Karen wishes to open. There is nothing in this that specifies that these two objects should in fact be the same object: that aspect of the scenario is additional, and as such is specific to $E_3$. Our model shows that, without an external trigger to set the evolutionary process in motion, it is not possible, in a coded communication system, for a new, non-composite signal like this to emerge (principle 2, above). Yet here, not only does such a signal emerge, it does so immediately, and smoothly: there is no interruption of the normal flow of communication. Neither is this an instance of the emergence of communication by ritualization, in which a cue evolves into a signal (see principle 3, above) - because Karen's behaviour is not a cue. It is a signal from the moment of its production, and that is the point. As such, this is a clear exception to the general constraints described previously. In sum, the existence of ostensive communication makes it possible for a species to overcome the constraints, described above, that otherwise make the emergence of combinatorial communication unlikely.

There is, then, an important sense in which Karen's twist signal contrasts with, say, bacterial quorum sensing, or the putty-nosed monkey case. Systems like that, which appear to be cases of coded communication, are in a sense not really combinatorial at all. After all, there is no 'combining' going on. There is really just a third holistic signal, which happens to be comprised of

the same pieces as other existing holistic signals. Indeed, the most recent experimental results suggest that the putty-nosed monkeys interpret the 'combinatorial' pyow-hack calls in exactly this idiomatic way, rather than as the product of two component parts of meaning [30]. In contrast, the ostensive creation of new composite signals is clearly combinatorial: the meaning of the new, composite signal is in part (but only in part) a function of the meanings of the component pieces.

It is presently unclear whether any other species uses ostensive communication. The precise psychological mechanisms necessary are cognitively complex, and so it is quite possible that it is uniquely human [31-33]. Certainly, this would be consistent with the argument we have developed in this paper, and there is presently no convincing evidence that any other species communicates ostensively [31,33]. However, this remains, at least for now, an open empirical question. (Note that ostensive communication is not the same thing as intentional communication, which some other species certainly do use.)

## 8. Conclusion

Previous models of the emergence of combinatorial communication were focused upon the following question: under what circumstances are composite signals advantageous, in comparison with holistic signals? Our model in this paper addresses a different question: by what processes can composite signals emerge? To do this, we explicitly modeled the possibility that no communication might take place: this is why our model includes the default states $E_0$, $A_0$ and $R_0$, which are absent from other models. Our results show that combinatorial communication is rare in nature because the interdependence of signals and responses constrains the ways by which communication systems emerge, with the effect that novel signals will tend to be holistic rather than composite. However, this constraint can be bypassed if the communication system in question is ostensive - and this type of communication is likely unique to humans. Unlike other proposals (see *Introduction*), this explanation is consistent with all the empirical facts: it explains both why combinatorial

communication is generally rare in the natural world, and why there is a single, extreme exception to this trend.

**Acknowledgements**

**References**

1.  Arnold, K. & Zuberbühler, K. 2006 Semantic combinations in primate calls. *Nature, 441*, 303.

2.  Arnold, K. & Zuberbühler, K. 2008 Meaningful call combinations in a non-human primate. *Curr Biol, 18*, R202-R203.

3.  Nowak, M. A., Plotkin, J. B. & Jansen, V. A. 2000 The evolution of syntactic communication. *Nature, 404*, 495-498.

4.  Fitch, W. T. 2010 *The Evolution of Language*. Cambridge: Cambridge University Press.

5.  Hurford, J. R. 2011 *The Origins of Grammar*. Oxford: Oxford University Press.

6.  Scott-Phillips, T. C., Gurney, J., Ivens, A., Diggle, S. P., & Popat, R. (in revision) The extent of combinatorial communication in nature.

7.  Zuberbühler, K. 2002 A syntactic rule in forest monkey communication. *Anim Behav, 63*, 293-299.

8.  Ouattara, K., Lemasson, A. & Zuberbühler, K. 2009 Campbell's monkeys concatenate vocalizations into context-specific call sequences. *P Nat Acad Sci, 106*, 22026-22031.

9.  Nowak, M. A. & Krakauer, D. C. 1999 The evolution of language. *P Nat Acad Sci, 96*, 8028-8033.

10. Hobaiter, C. & Byrne, R. W. (2011). The gestural repertoire of the wild chimpanzee. *Anim Cog*, 14(5), 745-767.

11. Lachmann, M.,& Bergstrom, C. T. 2004 The disadvantage of combinatorial communication. *P Roy Soc Lond B, 271*, 2337-2343.

12. Ridley, M. 2009 *Evolution*. Oxford: Wiley.

13. Bradbury, J. W. & Vehrencamp, S. L. 2000 Economic models of animal communication. *Anim Behav, 59*, 259-268.

14. Scott-Phillips, T. C., Blythe, R. A., Gardner, A. & West, S. A. 2012 How do communication systems emerge? *P Roy Soc Lond, B, 279*, 1943-1949.

15. Maynard Smith, J. & Harper, D. G. C. 2003 *Animal Signals*. Oxford: Oxford University Press.

16. Bradbury, J. W. & Vehrencamp, S. L. 1998 *Principles of Animal Communication*. Sunderland, MA: Sinauer Associates, Inc.

17. Scott-Phillips, T. C. 2008 Defining biological communication. *J Evo Biol, 21*, 387-395.

18. Thornhill, R. 1976 Sexual selection and nuptial feeding in *Bittacus apicalis* (Insecta: Mecoptera). *Am Nat, 110*, 529–548.

19. Searcy, W. A. & Nowicki, S. 2007 *The Evolution of Animal Communication*. Princeton, NJ: Princeton University Press.

20. Donaldson, M. C., Lachmann, M. & Bergstrom, C. T. 2007 The evolution of functionally referential meaning in a structured world. *J Theo Biol, 246*, 225-233.

21. Jäger, G. 2008 Evolutionary stability conditions for signaling games with costly signals. *J Theo Biol, 253*, 131-141.

22. Nowak, M. A. 2006 *Evolutionary Dynamics: Exploring the Equations of Life*. Harvard, NJ: Belknap Press.

23. Hockett, C. F. 1969 The origin of speech. *Sci Am, 203*, 88-111.

24. Ladd, D. R. 2012 What is duality of patterning, anyway? *Lang Cog, 4*, 261-273.

25. Chomsky, N. 1965 *Aspects of the Theory of Syntax*. Cambridge, MA: MIT Press.

26. Sperber, D. & Wilson, D. 1995 *Relevance: Communication and Cognition* (2nd ed.). Oxford: Blackwell.

27. Origgi, G. & Sperber, D. 2010 A pragmatic perspective on the evolution of language. In *The Evolution of Human Language: Biolinguistic Perspectives* (eds. R. K. Larson, V. Déprez, & H. Yamakido), pp. 124-132. Cambridge: Cambridge University Press.

28. Sperber, D. & Wilson, D. 2002 Pragmatics, modularity and mind-reading. *Mind Lang, 17*, 3–23.

29. Goldin-Meadow, S. 2005. *The Resilience of Language*. Abingdon: Psychology Press.

30. Arnold, K. & Zuberbühler, K. 2012 Call combinations in monkeys: Compositional or idiomatic expressions? *Brain Lang, 120*, 303-309.

31. Tomasello, M. 2008 *Origins of Human Communication*. Cambridge, MA: MIT Press.

32. Fitch, W. T., Huber, L. & Bugnyar, T. 2010 Social cognition and the evolution of language: Constructing cognitive phylogenies. *Neuron, 65*, 795-814.

33. Scott-Phillips, T. C. (under review) Non-human primate communication, pragmatics, and the origins of language.

**Figure legend**

**Figure 1**: *Combinatorial communication.*

In a combinatorial communication system, two (or more) *holistic* signals (*A* and *B* in this figure) are combined to form a third, *composite* signal (*A* + *B*), which has a different effect (*Z*) to the sum of the two individual signals (*X* + *Y*). This figure illustrates the simplest combinatorial communication system possible. Applied to the putty-nosed monkey system, the symbols in this figure are: *a* = presence of eagles; *b* = presence of leopards; *c* = absence of food; *A* = 'pyow'; *B* = 'hack' call; *C* = *A* + *B* = 'pyow-hack'; *X* = climb down; *Y* = climb up; *Z* ≠ *X* + *Y* = move to a new location. Combinatorial communication is rare in nature: many systems have a signal *C* = *A* + *B* with an effect *Z* = *X* + *Y*; very few have a signal *C* = *A* + *B* with an effect *Z* ≠ *X* + *Y*.

|  | state of world | signal | effect |
|---|---|---|---|
| **non-combinatorial communication** | $a$ → | $A$ → | $X$ |
|  | $b$ → | $B$ → | $Y$ |
|  | $c \neq a + b$ → | $C \neq A + B$ → | $Z \neq X + Y$ |
| **combinatorial communication** | $a$ → | $A$ → | $X$ |
|  | $b$ → | $B$ → | $Y$ |
|  | $c \neq a + b$ → | $C = A + B$ → | $Z \neq X + Y$ |



\title{Why is combinatorial communication rare in the natural world, and why is language an exception to this trend?\\[3ex]
Supporting Information}

\author{Thomas~C.~Scott-Phillips\footnote{Evolutionary Anthropology Research Group, Department of Anthropology, Durham University, Dawson Building, South Road, Durham, DH1 3LE, UK}~ and Richard~A.~Blythe\footnote{SUPA, School of Physics and Astronomy, University of Edinburgh, James Clerk Maxwell Building, Mayfield Road, Edinburgh EH9 3JZ, UK}}

\date{June 11th 2013}

\maketitle

\newpage
\section{Replicator equations for composite signalling strategies}

In evolutionary game theory, the dynamics of a strategy $S$ in a population is governed by the replicator equation \cite{now06}
\begin{equation}
\frac{{\rm d} x(S)}{{\rm d} t} = x(S) \left[ f(S; \{ x\} ) - \sum_{S'} x(S') f(S'; \{ x \}) \right]
\end{equation}
where $x(S)$ is the frequency of the strategy $S$ in the population, $f(S; \{ x\})$ is the growth rate of agents with that strategy \emph{given} the frequencies of all other strategies in the population, and the sum is over the set of strategies $S'$ that are in direct competition with $S$ (including $S$ itself).

In a model of communication, there are many different sets of competing strategies.  First, the action, $A$, performed by an agent, $\alpha$, can vary according to the state of the environment, $E$.  We assume that a single action $A_{\alpha}(E)$ is performed by agent $\alpha$ whenever the environment is in state $E$: previous work on communication \cite{don07,jag08} has shown that probabilistic strategies (where more than one action might be performed in a given environment) are not evolutionarily stable. Thus, if $\psi(A,E)$ is the frequency of agents that perform action $A$ in environment $E$, we have the replicator equation
\begin{equation}
\label{psidot}
\frac{ \rm d}{{\rm d} t} \psi(A,E) = \psi(A,E) \left[ u(A,E) - \sum_{A' \in {\cal A}} \psi(A',E) u(A',E) \right]
\end{equation}
where the sum is over all elements of the set, ${\cal A}$, of all possible actions and their composites. $u(A,E)$ is the growth rate of the action strategy $E\to A$ and depends on the frequencies of all other strategies in the population.  In the model of composite communication described in the main text, we assume that the action performed in an composite environment $E_1 \circ E_2$ is the composite of the actions performed in the environments separately, i.e., $A_{\alpha}(E_1) \circ A_{\alpha}(E_2)$ for agent $\alpha$.  This means that the frequencies $\psi(A,E)$ are explicitly defined for those environmental states $E$ that are not composites.

The situation for reactions is directly analogous: for each action $A$ that is performed, an agent can perform a single reaction $R$: for each action, the different possible reactions compete with one another. Hence we have
\begin{equation}
\label{phidot}
\frac{ \rm d}{{\rm d} t} \phi(R,A) = \phi(R,A) \left[ v(R,A) - \sum_{R' \in {\cal R}} \phi(R',A) v(R,A) \right]
\end{equation}
where here $\phi(R,A)$ is the frequency of the reaction strategy $A\to R$ in the population, ${\cal R}$ is the set of all possible reactions and $v(R,A)$ is the growth rate of that strategy which again depends on the frequencies of all other strategies in the population.

The crucial question, then, is what form the fitnesses $u(A,E)$ and $v(R,A)$ should take. First, we associate the growth rate $s(R|E)$ with reaction $R$ in environment $E$ (which may include composite environments). We also assign a cost $c_{\alpha}$ to an agent for maintaining a given set of signalling strategies. For simplicity, we assume that each non-default action strategy (i.e., environment $E$ for which $A_{\alpha}(E) \ne A_0$) costs an amount $\chi$ and that each non-default reaction strategy (i.e., action $A$ for which $R_{\alpha}(A)\ne R_0$) costs an amount $\eta$.  We further assume the state $E$ is present a fraction $f(E)$ of the time.  Finally, if $q_\alpha(R,E)$ is the probability that agent $\alpha$ performs the reaction $R$ in environment $E$ as a consequence of some other agent performing an appropriate action, the mean rate of offspring production (fitness) of that individual is
\begin{equation}
s_{\alpha} = \sum_{E,R} f(E) s(R|E) q_{\alpha}(R,E) - c_\alpha \;.
\end{equation}
The fitnesses $u(A,E)$ and $v(R,A)$ are then obtained by averaging over the set of individuals who have the specific strategy $E\to A$ or $A \to R$.

There are two approaches that can be taken to calculate these fitnesses that lead to exactly the same outcome.  The first is to assume, as in \cite{don07,jag08}, that all agents can observe the behaviour of all other agents (i.e., the population is spatially well-mixed) and that signallers receive the same payoff, $s(R|E)$, as the agents who respond to their signals.  The second is instead to assume that agents can only observe the behaviour of conspecifics (i.e., those with the same action and reaction strategies), thereby leading to an indirect benefit for actors who also perform the corresponding reaction.  In order that different strategies may compete with one another, we further require in this case that recombination is efficient, so that the fraction of individuals with two particular strategies is given by the product of their individual frequencies.  Below, we show that these two approaches---which amount to different ways to ensure the stability of cooperative behaviour in the population---lead to the same dynamics.

\subsection{Direct benefit to both signaller and receiver in a spatially-mixed population}

We consider first the case where an agent can observe the behaviour of all other agents, and the payoff for performing a reaction $R$ in environment $E$ is passed on from the reactor to actor.

The easiest fitness to evaluate is $v(R,A)$, i.e., that for a reaction strategy $A \to R$. This is because the payoff for performing a reaction is direct.  There are three contributions to this quantity.  The first comes from actors who perform $A$ in a (non-composite) environmental state $E$.  These actors make up a proportion $\psi(A,E)$ of the population, so under the assumption that agents are spatially-mixed, the probability $q(R,E)$ that a reactor interacts with such an actor is $\psi(A,E)$.  This first contribution is then simply
\[ v^{(1)}(R, A) = \sum_{E} f(E) s(R|E) \psi(A,E) \;. \]
Now we have to take into account that, if $A$ is a composite action $A_1\circ A_2$, it may be performed in the composite environment $E_1\circ E_2$.  In a spatially well-mixed population, the probability that a randomly chosen actor behaves in this way is $\phi(A_1,E_1) \phi(A_2,E_2)$.  To decide if a given combination of actions $A_1\circ A_2$ is equivalent to $A$, we employ the Kronecker delta symbol $\delta(A_1\circ A_2, A)$, which equals $1$ if the two arguments are equal, and zero otherwise.  Then, we find
\[  v^{(2)}(R, A) = \sum_{\langle E_1,E_2 \rangle} f(E_1 \circ E_2) s(R | E_1 \circ E_2) \sum_{A_1,A_2} \psi(A_1,E_1) \psi(A_2,E_2) \delta(A_1\circ A_2, A) \]
where $\langle E_1,E_2 \rangle$ denotes a sum over distinct pairs of non-default, non-composite environmental states.
The final contribution to the fitness comes from the fact that maintaining each non-default reaction decreases the payoff by an amount $\eta$, no matter what behaviour the agent actually engages in.  Since only fitness differences matter, we can equally ascribe a fitness benefit to the default reaction of $\eta$, again by using the Kronecker delta symbol:
\[ v^{(3)}(R, A) = \eta \delta(R, R_0) \;. \]
Adding these three terms together gives the expression quoted in the main text.

We now turn to the fitness $u(A,E)$ of the action strategy $E\to A$.  To do this we need to identify the mean growth rate of agents employing the strategies $A\to R$ for fixed $A$ but variable $R$.  Again, this has three contributions.  First, the probability $q(R,E)$ that a randomly chosen reactor exhibits the reaction $R$ to the action $A$ in environment $E$ is $\phi(R,A)$.  Hence, the first contribution to the fitness is
\[ u^{(1)}(A,E) = f(E) \sum_{R} \phi(R,A) s(R|E) \;. \]
The second contribution comes from the case where the environmental state $E$ co-occurs with some other state $E'$ (which is distinct from $E'$ and $E_0$).  In the composite state $E\circ E'$, the probability that the composite action $A\circ A'$ is observed by a randomly-chosen reactor is $\psi(A',E')$, given that action $A$ is already performed by some actor.  The probability that reactor also performs the reaction $R$ to $A\circ A'$ is $\phi(R,A\circ A')$.  Hence, the second contribution to the fitness is
\begin{equation}
 u^{(2)}(A,E) = \sum_{E' \ne E_0, E} f(E \circ E') \sum_{R} \sum_{A'} \phi(R,A \circ A') \psi(A', E') s(R|E \circ E') \;.
\end{equation}
Finally, we can assign a fitness advantage to the default strategy $A_0$ via
\[ u^{(3)}(A,E) = \chi \delta(A, A_0) \;. \]
Again, summing these three contributions together we obtain the expression for $u(A,E)$ given in the main text.

\subsection{Indirect benefit through kin discrimination with random mating}

The foregoing expressions for the fitnesses were obtained by using the fact that, when an actor or reactor is chosen at random from the population, the probability that it has a given strategy $E\to A$ or $A\to R$ is just given by the frequency of that strategy in the population, $\psi(A,E)$ or $\phi(R,A)$.  This is appropriate when agents are well mixed in space.  An alternative approach is to assume that agents interact only with their conspecifics.  Then, when considering an agent as an actor with the strategy $A\to E$, for example, and asking whether a reactor that agent interacts with exhibits the strategy $A\to R$, this is equivalent to asking whether that same actor also has the strategy $A\to R$.  In principle, strong correlations could build up between different strategies.  However, if we assume that some mating process acts so that offspring acquire random combinations of parents' strategies, and that this process acts sufficiently quickly that it reaches equilibrium on the timescale of the growth dynamics, then the probability that an agent has the strategy $A\to R$, say, is $\phi(R,A)$ no matter what other strategies it may possess.  Thus, asking questions about a single agent in this picture is equivalent to asking those same questions about randomly-chosen agents in the previous section.  Hence, the fitnesses that arise from this approach are exactly equivalent.  It is possible to show this more formally by deriving the replicator dynamics from first principles using, e.g., the Price equation \cite{pri70,pri72} as a starting point.

\subsection{Conditions for evolutionary stability}

Evolutionarily stable strategies are found by identifying stable fixed points of the replicator equations (\ref{psidot}) and (\ref{phidot}).  To obtain vanishing right-hand sides of these equations, we must have that $\psi(A,E)=0$ for all but one action $A$ in each environment $E$, and that $\phi(R,A)=0$ for all but one reaction $R$ to each action $A$.  Thus, only homogeneous populations are fixed points of the replicator equations.  The only exception to this is when multiple fitnesses have the same value: then one has neutral stability in mixed populations. This will rarely be the case in situations of interest to us: even then, stochastic contributions (not considered here) will tend to lead to a homogeneous population.  Hence, only homogeneous populations can be evolutionarily stable, as stated in the main text.

To determine whether a particular homogeneous population is evolutionarily stable, we need to examine the behaviour of deviations away from the corresponding fixed point in (\ref{psidot}) and (\ref{phidot}).  Ultimately, we find that the requirement for stability (i.e., that the Hessian matrix evaluated at a fixed point has negative eigenvalues \cite{hof98}) is satisfied only if 
\begin{align}
u(A',E) < u(A(E),E) &\quad \forall A' \ne A(E) \quad\mbox{in each environment $E$} \\
v(R',A) < v(R(A),A) &\quad \forall R' \ne R(A) \quad\mbox{for each action $A$} \;,
\end{align}
where $A(E)$ and $R(A)$ specify the actions and reactions performed by all agents in the homogeneous population of interest.  When one of the conditions stated above does not hold, the population is vulnerable to an instability.  For example, at a fixed point where the population has the rule $E \to A_1$, it would be vulnerable to reverting to the default action $E \to A_0$, if $u(A_0, E) > u(A_1, E)$ when evaluated at the fixed point corresponding to homogeneous use of the rule $E\to A_1$.  The principles stated in the main text are obtained by investigating the situations under which different homogeneous fixed points are stable or, if they are unstable, what the nature of the instabilities are.

\section{Case study of the principles for the emergence of combinatorial communication: Putty-nosed monkeys}

We illustrate the general principles for the emergence of combinatorial communication set out in the main text with the concrete example of the putty-nosed monkey's communication system.  To recap, there are three basic environmental states that are relevant to communication: ${\bf L}$, where leopards are present; ${\bf E}$, where eagles are present; and ${\bf X}$, in which food is scarce.  We assume that the optimal behaviour in these environmental states is ${\bf U}$, to move up, ${\bf D}$, to move down, and ${\bf F}$, to flee, respectively.  We also assume that the only composite state that may exist is presence of both predators ${\bf L}\circ{\bf E}$, and that the optimal response in this environment is to flee (${\bf F}$), since moving away from one of the predators (${\bf U}$ or ${\bf D}$) will inevitably entail moving towards from one of the predators.

Taking into account the default environment $E_0$ and the default reaction $R_0$, we find that even this simple model has twenty different growth rates $s(R|E)$.  To keep the parameters to a manageable number, we introduce a set of costs $\alpha$, $\beta$, $\gamma$ and $\epsilon$ which relate to the presence of a predator, the absence of food, moving away from a predator and fleeing respectively, and a benefit $\delta$ for moving away from a predator.  Combining these costs additively leads to the set of growth rates specified in Table~\ref{tab:rates}.  In addition to these growth rates, we must also specify the costs $\eta$ and $\chi$ for maintaining components of a communication system, and the frequencies $f(E)$ with which the various environmental states are present.  Again, for simplicity, we assume that leopards and eagles are equally frequent, $f({\bf L})=f({\bf E})$.

To complete the definition of the model, we specify three actions ${\bf R}, {\bf G}$ and ${\bf B}$ that we construe as different colours (rather than sounds like `pyow' and 'hack'), and the single composite action ${\bf R} \circ {\bf G}$. 

\begin{table}
\begin{center}
\begin{tabular}{c|c|c|c|c}
& ${\bf R_0}$ & ${\bf U}$ & ${\bf D}$ & ${\bf F}$ \\\hline
${\bf E_0}$ & $0$ & $-\gamma$ &  $-\gamma$ &  $-\epsilon$ \\
${\bf L}$ & $-\alpha$ & $-\alpha-\gamma+\delta$ & $-\alpha-\gamma$ & $-\epsilon$ \\
${\bf E}$ & $-\alpha$ & $-\alpha-\gamma$ &$-\alpha-\gamma+\delta$ & $-\epsilon$ \\
${\bf X}$ & $-\beta$ & $-\beta-\gamma$ & $-\beta-\gamma$ & $-\epsilon$ \\
${\bf L}\circ{\bf E}$ & $- 2\alpha$ & $- 2\alpha-\gamma+\delta$ & $- 2\alpha-\gamma+\delta$ & $-\epsilon$ 
\end{tabular}
\end{center}
\caption{\label{tab:rates} Growth rates $s(R|E)$ as a function of the behaviour $R \in \{ R_0, {\bf U}, {\bf D}, {\bf F} \}$ in each environmental state $E \in \{ E_0, {\bf L}, {\bf E}, {\bf X}, {\bf L}\circ {\bf E} \}$. The idea is that there is a penalty $\alpha$ when one predator is present, and $2\alpha$ when both are present; a cost $\beta$ if food is scarce; a cost of moving away from predator $\gamma$; a benefit $\delta$ if the movement away from a predator leads to a greater chance of survival; and a cost of $\epsilon$ for fleeing.  It is assumed that all the costs and benefits are cumulative, except for fleeing which (by taking the agent to a completely new location) is taken to be independent of the environmental state.}
\end{table}

We now demonstrate how principles 2 and 3 stated in the main text allow us to understand the constraints on reaching the actual communication system exhibited by putty-nosed monkeys (i.e., the signals ${\bf L} \to {\bf R} \to {\bf U}$, ${\bf E} \to {\bf G} \to {\bf U}$ and ${\bf X} \to {\bf R}\circ{\bf G} \to {\bf F}$) from a simpler system.  Since one needs at least two signals to create a composite signal, we consider starting points whereby two signals have separately evolved.  There are two distinct starting points: (A) one in which one of the predator signals is present alongside the food signal, i.e., ${\bf L} \to {\bf R} \to {\bf U}$ (or ${\bf E} \to {\bf G} \to {\bf U}$ which is equivalent, due to the symmetry in the predators we have built into this model) and ${\bf X} \to {\bf B} \to {\bf F}$; and (B) one in which both predator signals are present, i.e., ${\bf L} \to {\bf R} \to {\bf U}$ and ${\bf E} \to {\bf G} \to {\bf U}$.  Note that in the former case, there was no option other than to use the non-composite action ${\bf B}$ to act as a cue for the absence of food.

\begin{figure}

\hspace*{1cm}
\includegraphics[bb=50 50 1000 500,width=12.2cm]{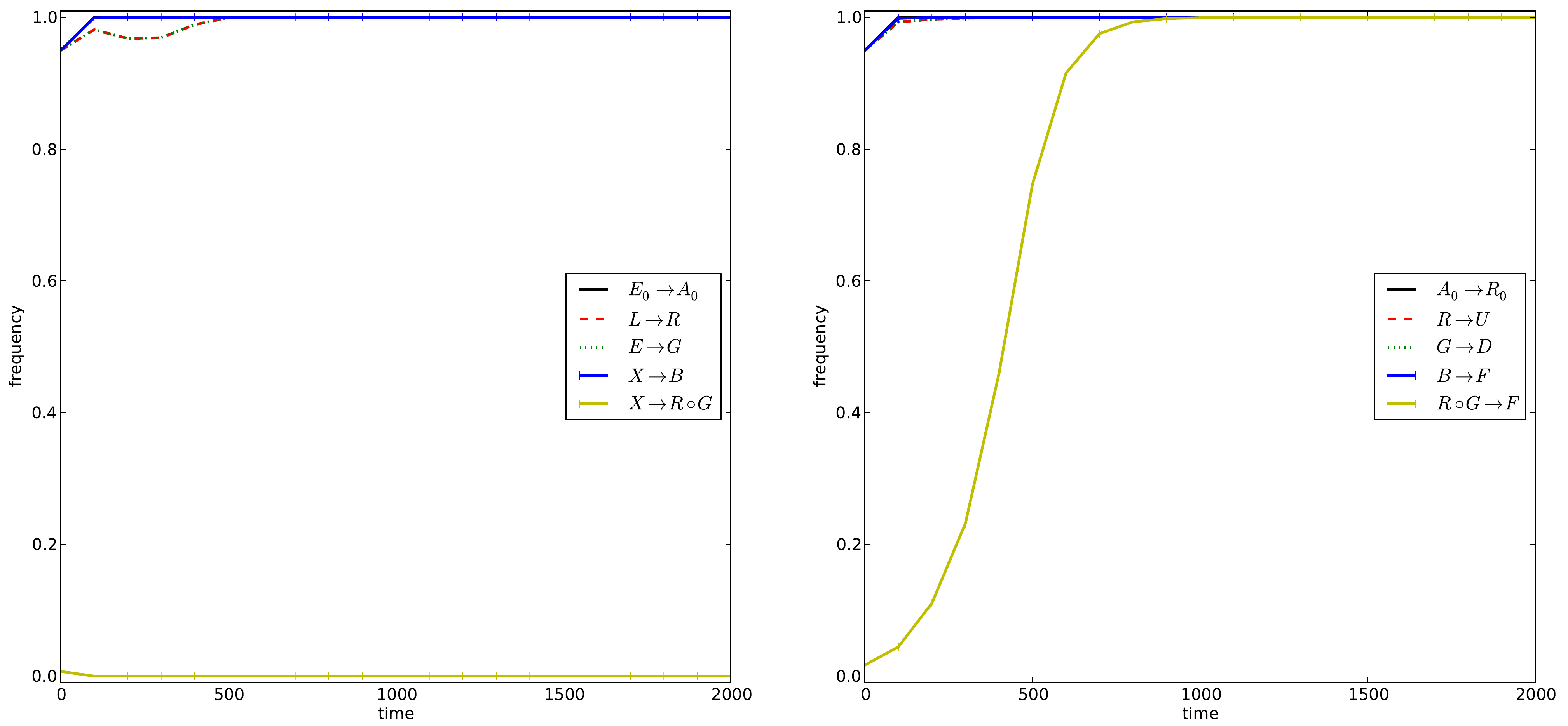}
\medskip

\caption{\label{fig:ritonly} Ritualisation of a pseudo-composite signal from a system of three non-composite signals in the absence of an external trigger.  The left panel shows the frequencies $\psi(A,E)$ of relevant action strategies; the right panel shows the frequencies of $\phi(R,A)$ of relevant reaction strategies. These were obtained obtained by direct numerical integration of the system (\ref{psidot}) and (\ref{phidot}) with the set of model parameters given in Table~\ref{tab:values}.}
\end{figure}

Principle 2 in the main text states that given starting point (A), the only way that a new signal can be added is by some external trigger: if the system (A) is unstable, it can be unstable only to losing existing actions or reactions, rather than by adding a new action or reaction, since these necessarily incur a cost $\eta$ or $\chi$ respectively to no benefit.  The only way a more complex system can be constructed in this case is the additional of a new signal by an external trigger. The only signal absent from (A) is the remaining predator signal.  This leads to a system in which all three non-composite environmental states map to distinct non-composite actions, each of which yields a distinct reaction.  It is possible that once this second predator signal is added, ritualisation of the action ${\bf R}\circ{\bf G}\to {\bf F}$ may occur due to ${\bf R}\circ{\bf G}$ being performed in the composite state ${\bf L}\circ{\bf E}$, and because the flee reaction is optimal in the presence of both predators. This leads to the establishment of a \emph{pseudo-composite} signal within the classification scheme outlined in the main text.  For the set of model parameters given in Table~\ref{tab:values}, we find that this is exactly what happens: see Figure~\ref{fig:ritonly} which shows the results of direct numerical integration of the ODE system (\ref{psidot}) and (\ref{phidot}) under these conditions.

\begin{figure}
\hspace*{1cm}
\includegraphics[bb=50 50 1000 500,width=14.5cm]{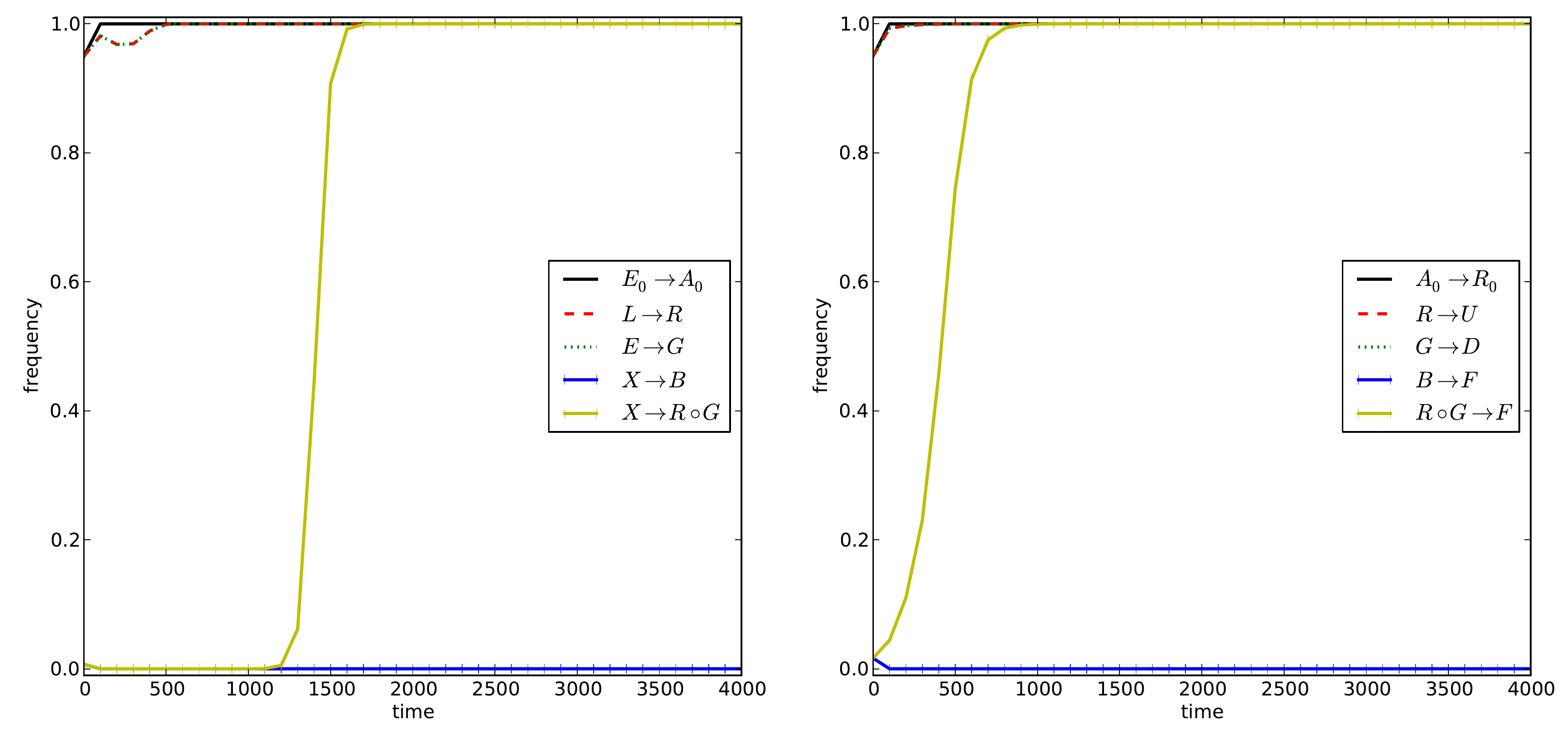}
\medskip

\caption{\label{fig:ritandsensmanip} Evolution of a fully-composite signal from a system of two non-composite signals in the absence of an external trigger.  This is achieved first by ritualisation of ${\bf R}\circ{\bf G} \to {\bf F}$ which creates a state that is then vulnerable to sensory manipulation of ${\bf X} \to {\bf R}\circ{\bf G}$. The left panel shows the frequencies $\psi(A,E)$ of relevant action strategies; the right panel shows the frequencies of $\phi(R,A)$ of relevant reaction strategies. These were obtained obtained by direct numerical integration of the system (\ref{psidot}) and (\ref{phidot}) with the set of model parameters given in Table~\ref{tab:values}.}
\end{figure}

Now we consider starting point (B).  Principle 3 in the main text states that this starting point may be vulnerable to the emergence of a composite signal without any external triggering.  As we further discuss in the main text, this is permitted when the system comprising three non-composite signals is unstable.  In particular, this is true for the combination of parameter values given in Table~\ref{tab:values}.  Figure~\ref{fig:ritandsensmanip} shows that ritualisation of ${\bf R}\circ{\bf G}\to {\bf F}$ is followed by sensory manipulation of ${\bf X} \to {\bf R}\circ{\bf G}$, generating the putty-nosed monkey's communication system without the need for an external trigger.  On the other hand, when the non-composite signalling system is stable, this pathway is suppressed and the starting point (B) is expected to be stable, even if the system with a fully-composite signal is also stable (and therefore, in principle, a possible endpoint of the dynamics).  Direct numerical integration of (\ref{psidot}) and (\ref{phidot}) under these conditions shows no change in strategy frequencies over time, indicating that the two-signal system (B) is stable in this case, and further signals can be added only by means of an external trigger.  As we argue in the main text, the most likely scenario is that the stable system of three non-composite signals will be reached through such a mechanism, rather than the system with a composite signal, even though that is also stable.

\begin{table}
\begin{center}
\begin{tabular}{c|c|c|c|c|c|c|c|c|c|c|c}
$f(E_0)$ & $f({\bf L})$ & $f({\bf E})$ & $f({\bf L}\circ{\bf E})$ & $f({\bf X})$ & $\alpha$ & $\beta$ & $\gamma$ & $\delta$ & $\epsilon$ & $\chi$ & $\eta$ \\\hline
0.38 & 0.2 & 0.2 & 0.07 & 0.15 & 1.5 & 2.6 & 1.0 & 1.3 & 2.1 & 0.05 & 0.05 \\
\end{tabular}
\end{center}
\caption{\label{tab:values} Parameters used in direct numerical integration of the system of ODEs (\ref{psidot}) and (\ref{phidot}).}
\end{table}

This simple example thus demonstrates two constraints on the trigger-free emergence of fully-composite signals: (i) in order for a composite action to be used as a cue for an unrelated environmental state, no other action may be in use as part of a signal for that state; and (ii) the system in which the unrelated environmental state is signalled by an non-composite action must in itself be unstable.


\newpage
\begin{thebibliography}{8}
\providecommand{\natexlab}[1]{#1}
\providecommand{\url}[1]{\texttt{#1}}
\expandafter\ifx\csname urlstyle\endcsname\relax
  \providecommand{\doi}[1]{doi: #1}\else
  \providecommand{\doi}{doi: \begingroup \urlstyle{rm}\Url}\fi

\bibitem[Nowak(2006)]{now06}
M~A Nowak.
\newblock \emph{Evolutionary dynamics: Exploring the equations of life}.
\newblock Belknap Press, Harvard, 2006.

\bibitem[Donaldson et~al.(2007)Donaldson, Lachmann, and Bergstrom]{don07}
M~C Donaldson, M~Lachmann, and C~T Bergstrom.
\newblock The evolution of functionally referential meaning in a structured
  world.
\newblock \emph{J. Theor. Biol.}, 246:\penalty0 225--33, 2007.

\bibitem[J\"{a}ger(2008)]{jag08}
G~J\"{a}ger.
\newblock Evolutionary stability conditions for signaling games with costly
  signals.
\newblock \emph{J. Theor. Biol.}, 253:\penalty0 131--41, 2008.

\bibitem[Price(1970)]{pri70}
G~R Price.
\newblock Selection and covariance.
\newblock \emph{Nature}, 227:\penalty0 520, 1970.

\bibitem[Price(1972)]{pri72}
G~R Price.
\newblock Extension of covariance mathematics.
\newblock \emph{Ann. Hum. Genet., London}, 35:\penalty0 485--90, 1972.

\bibitem[Hofbauer and Sigmund(1998)]{hof98}
J~Hofbauer and K~Sigmund.
\newblock \emph{Evolutionary games and population dynamics}.
\newblock Cambridge University Press, Cambridge, UK, 1998.


\end{thebibliography}
\end{document}